\documentstyle[12pt]{article}
\date{}
\begin{document}

\title{Two-Pion Exchange Interaction Between Constituent Quarks}

\author{D. O. Riska$^1$ and G. E. Brown$^2$}
\maketitle

\centerline{\it $^1$Department of Physics, 00014 University of
Helsinki, Finland}

\centerline{\it $^2$Department of Physics, State University of New York,
Stony Brook, NY 11794, USA}

\vspace{1cm}

\centerline{\bf Abstract}
\vspace{0.5cm}

The two--pion exchange interaction between constituent quarks is shown
to enhance the effect of the the isospin dependent spin--spin
component of the one--pion exchange interaction, and to cancel
out its tensor component. It therefore provides a partial
explanation for the phenomenological observation that 
the hyperfine interaction between constituent quarks is well
described by a flavor dependent spin--spin
interaction, which is attractive at short and repulsive at
long range. The spin--orbit component of the
two--pion exchange interaction is stronger than and 
has the opposite sign from that associated with the linear
confining interaction
in the $P-$shell multiplets.

\vspace{1cm}

\centerline{PACS: 12.39.Pn,12.39.-x,14.20.-c}

\vspace{0.5cm}

\centerline{KEYWORDS: Quark-quark interaction, two--pion exchange}

\newpage

{\bf 1. Introduction}
\vspace{0.5cm}

	The instanton liquid model of the vacuum 
\cite{Shur,Nowak} implies
pointlike quark--quark interactions, which when iterated in the
$t-$channel, admit a meson exchange interpretation of the
interaction between the constituent quarks that 
form the baryons \cite{GloVar}.
This description is supported by explicit QCD lattice calculations,
which suggest that the chromomagnetic interaction between
gluons is screened out at the 
length scales $\sim 0.3$ fm characteristic of the instanton fluctuations 
\cite{Negele1,Negele2,KFLiu}. Phenomenological investigation 
of the baryon spectrum,
as presently known, does indeed reveal that main features of the baryon
spectrum up to the spin--orbit splittings, may be described by a
flavor-spin dependent interaction, with a form similar to the short
range part of
the flavor--spin component of
$\pi,K$ and $\eta-$meson exchange interaction between
constituent quarks alone \cite{GloRis,Varga}. 
Even so there remain a number of gaps in this
description.

	To these gaps belongs the fact that, while the flavor--spin
component of the pseudoscalar meson exchange interaction
has the spectroscopically desired features, the corresponding
tensor component by itself would cause small but empirically 
contraindicated
spin--orbit splittings of the low lying negative parity multiplets in
the baryon spectrum. Another unclosed gap is the need for a
repulsive spin--orbit interaction component to cancel
the
attractive (Thomas term) spin--orbit interaction, which is associated 
with the linear
scalar confining interaction. 
The former one of these problems may be
solved qualitatively by invoking vector meson exchange interactions
in addition to the pseudoscalar meson exchange interaction between the
constituent quarks \cite{Wagenbrunn}. 
An alternate source of cancellation for the tensor interaction
is the tensor
component that is associated with the irreducible $\pi-$gluon exchange
interaction, 
which has significant strength even
when the gluon coupling to constituent quarks is screened at large
distances \cite{HelRis}. Invocation of such interaction
mechanisms
does however, inevitably, raise
the question of the relative importance of the related irreducible
two--pion exchange interaction. This issue is addressed
here by an 
explicit calculation of the two--pion exchange interaction between
constituent quarks.

	Before proceeding is should be noted that the latter spin--orbit
problem may be avoided phenomenologically by postulation
of a strong 
gluon exchange
term in the hyperfine interaction
\cite{Isgur}, although
only at the price of incorrect spectral ordering \cite{GloRis} and 
conflict
with the QCD lattice result
that the gluonic part of the chromomagnetic hyperfine
interaction should be very weak at the relevant 
length scales \cite{Negele1,Negele2,KFLiu}.

	Turning then to the two--pion exchange interaction between
constituent quarks we note that
the pseudovector pion--quark coupling constant $f_{\pi qq}$ is smaller
by a factor $\sim$ 3/5 than the corresponding pion--nucleon coupling constant
$f_{\pi NN}\simeq 1$, while the constituent quark mass is about 
$\sim$ 1/3 of the
nucleon mass. Therefore the {\it a priori} expectation should be that the
strength of the two--pion exchange interaction between constituent
quarks should -- on the average -- be about as strong as the 
corresponding
two--pion exchange interaction between nucleons, as the scaling factor
implied by the coupling constant and the mass ratios is
$(3/5)^4(1/3)^{-2}\simeq 1.2$. We find this indeed to be the case, and
that for quark separations smaller than $\simeq$ 0.6 fm,
which is the scale relevant for baryon structure, the components of
the two pion exchange interaction do dominate over those of the one--pion
exchange interaction.

The calculational tools required to construct a properly defined
irreducible two--pion exchange interaction are available in the
literature, having originally been developed for the two--pion exchange
interaction between nucleons \cite{Partovi,Chemtob,Brown}. These 
methods
may, with but minor modifications, be applied to the interaction between
constituent quarks,
provided that the effect of quark confinement
on the quark propagators may be approximately incorporated
in the effective constituent quark mass. 
The interaction is constructed from the
Poincar\'e invariant 4--point function, which is used to define
a quasipotential for the covariant Blankenbecler--Sugar equation
\cite{Logun,Blanken}, from which the iterated one--pion exchange
interaction is subtracted. The covariance requirement is essential
in the case of light constituent quarks, which have velocities close
to that of light, when confined within the baryons.

The spontaneous breaking of 
chiral symmetry requires the pion--quark coupling to have 
pseudovector form. The pions are assumed to decouple from the
constituent quarks at the chiral symmetry restoration scale
$\Lambda_\chi \sim$ 1 GeV, and therefore no ultraviolet divergence
appears, although the calculated values of the potential 
at short range are sensitive to the way this cut--off is
implemented. While the calculation of the two-pion exchange loop
diagrams is straightforward -- if both ``lengthy and tedious'' --
those loop diagrams do not by themselves provide a realistic
estimate of the two--pion exchange interaction. This is strongly
influenced by the strong interaction between the exchanged
pions in the $S-$ and $P-$states in the $t-$channel, which
has to be taken into account. This may be done by separating out
the contributions of those to partial waves from the loop
amplitudes, and replacing them with amplitudes, which do
contain the main resonance contributions, in addition to
the Born terms \cite{Chemtob,Durso}.

	The main features of the two--pion exchange interaction may be
summarized as follows: (1) it has a strong spin and isospin
dependent central interaction, which adds to the corresponding
component of one--pion exchange, 
(2) its isospin dependent
tensor component in effect cancels the tensor component of
one--pion exchange, (3) its spin--orbit components combine
in isospin antisymmetric quark pair states to a repulsive
net spin--orbit interaction, which overwhelms the spin--orbit
component of the linear scalar confining interaction
in the $P-$shell multiplets.
These features conform with the phenomenological
observation that the baryon spectrum is well described
by a linear confining interaction in combination with a
flavor dependent spin--spin interaction, which is attractive
at short and repulsive at long range. 
The one--plus--two--pion exchange hyperfine interaction is attractive in
quark pair states with $T=S=0$, and repulsive in
quark pair states with $T=S=1$.

This paper falls into 6 sections. In section 2  
the
calculational method is described. The method for treating
the strong $\pi\pi$
interaction in the $J=0,1$ states in the $t-$channel
is described in section 3. The
numerical results are given in section 4. Section 5 contains
a discussion of the phenomenological implications and the
role of $\omega$ and gluon exchange between quarks.
Section 6 contains a summarizing
discussion.

\vspace{1cm}

{\bf 2. The two--pion exchange potential}\\

{\bf 2.1. The one--pion exchange interaction}\\
\vspace{0.5cm}

The spontaneously broken approximate chiral symmetry of QCD suggests
that to lowest order in $(p_\pi/m_\pi)$ the coupling of pions to
constituent quarks have the form 
$${\cal L}_{\pi qq}=-{1\over f_\pi}\partial_\mu\vec\phi_\pi\cdot \vec
A_\mu,\eqno(2.1)$$
where $\vec A_\mu=ig_A\bar q\gamma_\mu\gamma_5\vec \tau q/2$ is the
axial current of the quark and $f_\pi$ is the pion decay constant (93
MeV). The effective (pseudovector) pion--quark coupling constant is
then 
$$f_{\pi qq}={g_A\over 2}{m_\pi\over f_\pi}.\eqno(2.2)$$
With $g_A\simeq 0.87$ for the quark
\cite{Weinberg,Dicus}, this yields $f_{\pi qq}\simeq
0.65$, which is close to the static quark model value $f_{\pi
qq}=3/5 f_{\pi NN}\simeq 0.6$, where $f_{\pi NN}$ is the pseudovector
$\pi-$nucleon coupling constant.

The interaction (2.1) immediately yields the one--pion exchange
interaction 
$$K_\pi={f_{\pi qq}^2\over m_\pi^2}{(\gamma^{(1)}\cdot
k)\gamma_5^{(1)}(\gamma^{(2)}\cdot k)\gamma_5^{(2)}\over
m_\pi^2+k^2}f(k^2)\vec \tau^1\cdot \vec \tau^2.\eqno(2.3)$$
Here $f(k)$ is a cut--off function, the role of which is to 
switch off the
pion--quark coupling at the chiral symmetry restoration scale
$\Lambda_\chi$. We shall take this to have the simple monopole form 
$$f(k^2)={\Lambda^2_\chi-m_\pi^2\over \Lambda_\chi^2+k^2},\eqno(2.4)$$
where $\Lambda_\chi \simeq 1$ GeV.\\

\vspace{1cm}

{\bf 2.2. Definition of two--pion exchange interaction}
\vspace{0.5cm}

The $\pi-$quark coupling (2.1) leads in 4th order to the two--pion
exchange 4--point functions illustrated diagrammatically in Fig.1. For
the construction of the corresponding irreducible interaction the
iteration of the one pion exchange interaction (2.3) has to be
subtracted from the amplitude with uncrossed pion lines (Fig.1a). For
this purpose we employ the framework of the Blankenbecler--Sugar
equation \cite{Logun,Blanken}, which may be written formally as the
integral equation 
$$M=U+U\tilde{g}M,\eqno(2.5)$$
where $M$ is the 4--point function, $U$ the interaction 
(quasipotential)
and $\tilde{g}$ the 3--dimensional two--quark propagator
$$\tilde{g}(\vec k,W)=2\pi i\delta(k_0){\Lambda_+^{(1)}(\vec
k)\Lambda_+^{(2)}(-\vec k)\over E^2(\vec k)-W^2-i\epsilon}.\eqno(2.6)$$
Here $W$ is the total energy and $\Lambda_+^{(i)}(\vec k)\,\,(i=1,2)$ are
positive energy projection operators \cite{Partovi}. In (2.6)
$E^2(\vec k)=m^2+\vec k^2$, where $m$ is the constituent
quark mass. For small quark momenta the Blankenbecler--Sugar
equation adiabatically approaches the form of the Lippmann--Schwinger
equation, and therefore the conventional phenomenological
Hamiltonian approach to the quark model. 

The relation between the interaction $U$ and the 4--point function $M$
is defined by the equation 
$$U=K+K(G-\tilde{g})U,\eqno(2.7)$$
where $K$ is an irreducible interaction kernel and $G$ is the product
of two free fermion propagators for the quarks.

Given the one--pion exchange interaction $K_\pi$ (2.3),
perturbative solution of (2.5) leads to the following formal expressions
for the one-- ($U_\pi$) and two--pion exchange ($U_{\pi\pi}$)
components of the quasipotential: 
$$U_\pi=K_\pi,\eqno (2.8a)$$
$$U_{\pi\pi}=M_{\pi\pi}-K_\pi\tilde g K_\pi.\eqno (2.8b)$$
Here $M_{\pi\pi}$ is the full two--pion exchange 4--point
function.

The explicit expressions for the two--pion exchange interaction
$U_{\pi\pi}$ have been given in ref.\cite{Chemtob} for the 
case of the nucleon--nucleon interaction, and may
be applied to the case of the two--pion
exchange interaction between constituent quarks with only
minor modifications, which are described below. These
include replacement of the pion--nucleon coupling constant
by the appropriate pion--quark coupling constant, and
of the nucleon mass by the pion mass in all expressions.
The main modification is that required by introduction
of the cutoff factor $f(k^2)$ (2.3) for the pion--nucleon
coupling.

\vspace{1cm}

{\bf 2.3. Calculational Details}
\vspace{0.5cm}

The two--pion exchange 4--point function $M_{\pi\pi}$ is
calculated under the assumption that the external quarks
satisfy the Dirac equation. The amplitude may then be
decomposed into 5 linearly independent spin invariants,
the choice \cite{Amati}:
$$P_1=1, \quad P_2=i(\gamma^1\cdot P+\gamma^2\cdot N),$$
$$P_3=(i\gamma^1\cdot N)(i\gamma^2\cdot P), \quad
P_4=\gamma^1\cdot\gamma^2, \quad P_5=\gamma^1_5\gamma^2_5,
\eqno(2.9)$$
being particularly convenient. Here $P=(p_1+p'_1)/2$ and
$N=(p_2+p'_2)/2$, where $p_1, p_2$ and $p'_1, p'_2$
are the 4--momenta of the initial and final quarks,
respectively. 

The two--pion exchange amplitude then takes the form
$$M_{\pi\pi}=\sum_{j=1}^5[3p_j^+(s,t,u)+
2p_j^-(s,t,u)\vec\tau^1\cdot\vec\tau^2].\eqno(2.10)$$
Here $s,t$ and $u$ are the invariant 
variables defined as
$$s=-(p_1+p_2)^2,\quad t=-(p'_1-p_1)^2,
\quad u=-(p'_2-p_2)^2.\eqno(2.11)$$
The coefficient functions $p_j^\pm$ admit the spectral
representations
$$p_j^\pm(s,t,u)={1\over \pi}\int_{4m_\pi^2}^\infty
dt'{\rho_j^\pm(s,t')\mp(-)^j\rho_j^\pm(u,t')\over
t-t'}.\eqno(2.11)$$
The explicit expressions for the spectral weight functions
$\rho_j^\pm$ 
for the two--pion exchange amplitudes in Fig.1 
are given in ref.\cite{Chemtob}, and
may be employed in the present case, once the fermion
mass $m$ is interpreted as the quark mass, and the
pion--nucleon coupling constant $g$ is replaced
by the corresponding pion--quark coupling constant
$g=2(m_q/m_\pi)f_{\pi qq}$. The spectral functions
$\rho_j^{\pm}$ are formed as a sum of loop diagram
contributions and terms associated with the
$S-$ and $P-$wave $\pi\pi$ intermediate states as
described in section 3 below. This separation makes
it possible to take into account the interaction
between the exchanged pions as well as the constraints
of chiral symmetry for the $S-$wave amplitude as 
shown in \cite{Durso}.
 
The effect of including the cutoff factor $f(k^2)$ in the 
one--pion exchange
interaction (2.3) on the two--pion exchange amplitude
may be taken into account by
separation into partial fractions. In both
loop amplitudes in Fig. 1 there appears a product of two pion
propagators and two cut--off factors, which may
decomposed as 
$${f(k_1^2)\over m_\pi^2+k_1^2}{f(k_2^2)\over m_\pi^2+k_2^2}
={1\over m_\pi^2+k_1^2}{1\over m_\pi^2+k_2^2}$$
$$-{1\over \Lambda^2+k_1^2}{1\over m_\pi^2+k_2^2}
-{1\over m_\pi^2+k_1^2}{1\over \Lambda^2+k_2^2}
+{1\over \Lambda^2+k_1^2}{1\over \Lambda^2+k_2^2}.\eqno(2.12)$$
This implies that the two--pion exchange interaction will be
formed as a linear combination of 4 different terms, in which 
the masses of both the exchanged ``mesons'' take the values
$m_\pi$ and $\Lambda$ in turn.

The first and last terms on the r.h.s. of (2.12) are calculable 
directly using
the formulae in ref.\cite{Chemtob}. 
As the interaction is here only considered in the limit $W=2m$,
the two intermediate
terms in (2.12) may be calculated using the same formulae,
provided that (a) the lower limit in the dispersion integrals
(2.11) is replaced by $(m_\pi+\Lambda)^2$, and (b) the
variable combination $\sqrt{t'/4-m_\pi^2}=q$ (and its square) 
is replaced
everywhere where it occurs in the integrand
by the more general expression
$$q=\sqrt{ {t'^2-2t'(\Lambda^2+m_\pi^2)+(\Lambda^2-m_\pi^2)^2  
\over 4t'} } .\eqno(2.13)$$
The expressions  given in ref.\cite{Chemtob} for the
iterated one--pion exchange interaction term in (2.8b) may
also be employed when the cut--off factor is taken into
account as in (2.12). The modification required in those
expressions (when $W=2m$) is the replacement of the squared
pion mass $m_\pi^2$ everywhere in the corresponding integrands
by $(m_\pi^2+\Lambda^2)/2$ for the second and third terms
on the r.h.s. in 
(2.12) and by $\Lambda^2$ in the case of the fourth term.

\vspace{1cm}

{\bf 3. The $\pi\pi$ interaction in $S-$ and $P-$ waves}\\

{\bf 3.1 Partial wave projection in the $t-$channel}\\
\vspace{0.5cm}

The two--pion exchange loop amplitudes (Fig.1), when calculated
without account of the strong interaction between the two exchanged
pions,
do not provide a realistic description of the two--pion exchange
interaction. The strong interaction between the pions in the $S-$state
in the $t-$channel leads to a substantial enhancement of the
attractive part of the potential. In the $P-$state the $\pi\pi$
interaction so strong as to form the $\rho-$meson resonance. 
This has to be taken into account in order
to obtain a spin--orbit component with the phenomenologically required
sign.

The $\pi\pi$ correlations in the $S-$ and $P-$waves in the $t-$channel
may be taken into account by subtracting the $S-$ and $P-$wave
components in the $t-$channel from the two--pion exchange amplitude
$M_{\pi\pi}$, and then replacing those partial wave amplitudes from
the bare loop diagrams by realistic amplitudes, which contain
the effect of the $\pi\pi$ correlations.

Concretely this is implemented by decomposing the spectral functions
$\rho_j^{\pm}$ (2.11) in the form \cite{Chemtob}: 
$$\rho_j^\pm=d_j^\pm+b_j^\pm+c_j^\pm,\eqno(3.1)$$
where $d_j^\pm$ is the loop amplitude contribution (Fig.1), $b_j^\pm$
are the corresponding contributions from $S-$ and $P-$wave $\pi\pi$
intermediate states multiplied by --1 and $c_j^\pm$ are 
corresponding
amplitudes, which do take into account the $\pi\pi$ correlations. The
explicit expressions for the $S-$ and $P-$wave contributions $b_j^\pm$
are given in ref. \cite{Chemtob}, and may be employed here with the
modifications listed in section 2.3 above.

\vspace{1cm}

{\bf 3.2 The $\pi\pi$ $S-$wave interaction}
\vspace{0.5cm}

The $I=0$ $S-$wave $\pi\pi$ state only contributes to the amplitude
$c_1^+$ as defined in (3.1). This contribution may be expressed in
terms of the helicity amplitude $f_+^{(+)0}$ for $q\bar q\rightarrow
\pi\pi$ defined in analogy with that for \cite{Frazer} as
$$c_1^+(s,t)+c_1^+(u,t)={2\pi\over \sqrt{t'}}{q(t)\over
t'-4m^2}|f_+^{(+)0}(t)|^2.\eqno(3.2)$$
Here $q(t)$ is defined as in (2.13).

With the pseudovector coupling model (2.1) this amplitude is real and
takes the form \cite{Durso}
$$f_+^{(+)0}(t)_B={g^2\over 4\pi}\{{\chi^2\over m}-m(1-h\,arctan{1\over
h})\},\eqno(3.3)$$
where $\chi^2=m^2-t/4$ and $h=(q^2-t/4)/2q\chi$. Note that the 
function $h\,arctan(1/h)$ should be analytically continued to 
$(H/2)log[(H-1)/(H+1)]$, where $h=iH$ for $t>4m^2$, and that
$g=2(m_q/m_\pi)f_{\pi qq}$.

The $I=0$ $S-$wave $\pi\pi$ interaction is known to cause a large
enhancement of the magnitude of the helicity amplitude $f_+^{(+)0}(t)$
for $t-$values in the range $10-20m_\pi^2$ \cite{Nielsen}, and
possibly even to a wide resonance ($"\sigma"$), 
although the latter issue remains contentious 
\cite{Roos,Speth,Harada}. Such a
resonance contribution may be added to the Born term helicity
amplitude (3.3) as:
$$f_+^{(+)0}(t)=f_+^{(+)0}(t)_B-{ \chi^2\over 4\pi}
{g_{\sigma qq}g_{\sigma\pi\pi}\over
m^2_\sigma-t-i \gamma q(t)}.\eqno(3.4)$$
Here $m_\sigma$ 
represents the mass and of
the resonance, and $g_{\sigma qq}$ 
and $g_{\sigma\pi\pi}$ denote the 
$\sigma-$quark and $\sigma\pi\pi$ coupling strengths
respectively.
The parameter $\gamma$ is defined as
$\gamma=m_\sigma\Gamma/q(m_\sigma)$.

For the scalar meson resonance parameters we employ the suggested
values $m_\sigma=470$ MeV and $\Gamma=250$ MeV \cite{Roos}, from which
it follows that $\gamma=620$ MeV. To estimate the coupling constant
product $g_{\sigma qq}g_{\sigma \pi\pi}$ we fall back on the
$\sigma-$model. When applied to constituent quarks, this yields
$g_{\sigma qq}=m_q/f_\pi$ and $g_{\sigma \pi\pi}=m_\sigma^2/2f_\pi$.
With a constituent quark mass value of 340 MeV \cite{GloRis}
this then yields $g_{\sigma
qq}=3.65$ and $g_{\sigma\pi\pi}=2.5 m_\sigma =1.19$ GeV.

\vspace{1cm}

{\bf 3.3 The $\pi\pi$ $P-$wave interaction}
\vspace{0.5cm}

The $I=1$ $P-$wave $\pi\pi$ state contributes to the 
amplitudes $c_1^-,c_2^-$
and $c_4^-$ in (3.1). These contributions may be expressed in terms of
the helicity amplitudes $f_\pm^{(-)1}$ for $q\bar q\rightarrow \pi\pi$
as \cite{Chemtob}:
$$c_1^-(s,t)-c_1^-(u,t)=-{\pi\over
6}N(s-u)q^2(t)|\lambda^-|^2,\eqno(3.5a)$$
$$c_2^-(s,t)+c_2^-(u,t)=-{2\pi\over
3}Nq^2(t)Re\{\eta^{-*}\lambda^-\},\eqno(3.5b)$$
$$c_4^-(s,t)+c_4^-(u,t)=-{2\pi\over 3}N q^2(t)|\eta^-|^2.\eqno(3.5c)$$
Here $N=q(t)/32\pi^2\sqrt{t}$, and
$$\lambda^-={12\pi\over m^2-t/4}[f_+^{(-)1}(t)-{m\over
\sqrt{2}}f_-^{(-)1}(t)],$$
$$\eta^-=6\pi \sqrt{2}f_-^{(-)1}(t).\eqno(3.6)$$

The helicity amplitude combinations $\lambda^-$ and $\eta^-$ are
expressed as combinations of Born terms and a $\rho$--meson resonance
contribution:
$$\lambda^-=\lambda_B^-+\lambda^-_R,\quad
\eta^-=\eta^-_B+\eta^-_R.\eqno(3.7)$$
The Born term expressions are obtained as \cite{Durso}:
$$\lambda^-={3\pi mg^2\over 2q \chi^3}\{3h-(1+3h^2)arc tan {1\over
h}\},\eqno(3.8a)$$
$$\eta^-=-{3\pi g^2\over 2q\chi}\{h-(1+h^2)arc tan{1\over
h}\}.\eqno(3.8b)$$

The resonance contributions that correspond to the $\rho-$meson pole
are obtained as
$$\lambda^-_R=-{2\kappa g_{\rho qq}\over m}{g_{\rho \pi\pi}\over
m_\rho^2-t-i\gamma q^3(t)}.\eqno(3.9a)$$
$$\eta^-_R=2 g_{\rho qq}(1+\kappa){g_{\rho \pi\pi}\over
m_\rho^2-t-i\gamma q^3(t)}.\eqno(3.9b)$$
Here $m_\rho^2=0.59$ GeV and 
$\gamma = m_\rho\Gamma/q^3(m_\rho) = 2.5/$GeV
as determined from the empirical mass and width of the $\rho-$meson.

The quark model relation between the vector coupling constants of the
$\rho-$meson to quarks and to nucleons respectively as
obtained from the charge coupling term is
$$g_{\rho qq}=g_{\rho NN},\eqno(3.10)$$
whereas as obtained from the current coupling term it
is 
$$g_{\rho qq}(1+\kappa_{\rho qq})={3\over 5}({m_q\over m_N})
g_{\rho NN}(1+\kappa_{\rho NN}).\eqno(3.11)$$
With $g_{\rho\pi\pi}^2/4\pi\simeq 0.52$ the
first
relation gives the value $g_{\rho qq}=2.6$. 
Since for nucleons $\kappa_{\rho NN}=6.6$ \cite{Hohler},
it follows from the second relation that $\kappa_{\rho qq}=0.65$
when $m_q=340$ MeV. 

The $\rho \pi\pi$ coupling constant $g_{\rho\pi\pi}$ is conventionally
taken to be twice that of the $\rho-$nucleon coupling 
constant, but may
also be determined from the $\rho\rightarrow \pi\pi$ decay width. We
shall take it to have the value $g_{\rho\pi\pi}=2g_{\rho NN}=5.12$.
With these parameter values the model for the two--pion exchange
interaction that takes into account the strong $\pi\pi$ interactions
in the $S-$ and $P-$waves in the $t-$channel is completely specified.

\vspace{1cm}

{\bf 4. Numerical values of the two--pion exchange potential}

\vspace{0.5cm}

{\bf 4.1 The local part of the potential}

\vspace{0.5cm}

The most transparent way to illustrate the two--pion exchange
interaction between constituent quarks is to consider only the leading
local components in (asymptotic) series 
an expansion in $\vec p/m$ of the interaction.
Since the confined constituent quarks in the baryons have large
velocities, the leading local components of the interaction can
however at most give a qualitative description of the full
interaction. In view of the large uncertainties pertaining to the
coupling of pions to constituent quarks at short range, 
we shall here nevertheless be content with
a calculation of only the local components of the interaction in order
to obtain a qualitative understanding of the nature of the two--pion
exchange interaction between constituent quarks.

In the local approximation the spin invariants $P_j$ are reduced to
the potential operators 
$$\tilde{\Omega}_C=1,\quad \tilde{\Omega}_{LS}={1\over 2}i(\vec
\sigma^1+\vec \sigma^2)\vec p\,'\times \vec p,$$
$$\tilde{\Omega}_T=\vec \Delta^2(\vec \sigma^1\cdot \vec
\sigma^2)-3(\vec \sigma^1\cdot \vec \Delta)(\vec
\sigma^2\cdot \vec
\Delta),$$
$$\tilde{\Omega}_{SS}=\vec \sigma^1\cdot \vec \sigma^2,\eqno(4.1)$$
where $\vec \Delta=\vec p\,'-\vec p$. When expanded to order $\vec
p\,^2/m^2$ the general two--pion exchange interaction then in the
adiabatic limits takes the form
$$V=\sum_{\alpha}[\tilde{v}_\alpha^+(t)+\tilde{v}_\alpha^-(t)\vec
\tau^1\cdot \vec \tau^2]\tilde{\Omega}_\alpha,\eqno(4.2)$$
where $\alpha$ runs over the set (4.1),
and the potential coefficients
$v_\alpha^\pm(t)$ 
only depend on (invariant) momentum transfer. These
may be expressed as weighted integrals of the
spectral functions $\rho_j^\pm(t)$ and the corresponding weight
functions for the iterated one--pion exchange interaction. Explicit
expressions for these weight functions are 
given in ref. \cite{Chemtob}. 

Once the interaction potential is expressed in the form (4.2), it may
readily be Fourier transformed, and finally  -- to first order in
order $\vec p/m$ -- takes the form

$$V=\sum_{\alpha}[v_\alpha^+(r)+v_\alpha^- (r)\vec \tau^1\cdot
\vec\tau^2]\Omega_\alpha,\eqno(4.3)$$
where the set of spin operators $\{\Omega_\alpha\}$ is defined as
$$\Omega_C=1,\quad \Omega_{LS}=\vec S\cdot \vec L,$$
$$\Omega_T=S_{12},\quad \Omega_{SS}=\vec \sigma^1\cdot
\vec\sigma^2.\eqno(4.4)$$
The potential functions $v_\alpha^\pm(r)$ are then expressed as
integrals over Yukawa functions \cite{Chemtob}:
$$v_\alpha^\pm(r)=-{1\over
4\pi^2 r}\int_{4m_\pi^2}^{\infty}dt'\tilde{\rho}_\alpha(t')R_\alpha
(r\sqrt{t'})e^{-r\sqrt{t'}}.\eqno(4.5)$$
Here the weight functions $R_\alpha(r\sqrt{t'})$ are defined as
$$R_C(x)=1,\quad R_{LS}(x)=-{t'\over x}(1+{1\over x}),$$
$$R_T(x)=t'(1+{3\over x}+{3\over x^2}),\quad R_{SS}=-t'.\eqno(4.6)$$
The spectral functions $\tilde \rho_\alpha$ here are formed
as linear combinations of the spectral functions
$\rho_j^\pm$ (2.11). The explicit expressions for these
linear combinations are given in ref.\cite{Chemtob}\\

\vspace{1cm}

{\bf 4.2 Numerical results for the potential components}

\vspace{0.5cm}

The calculated components of the one and two--pion exchange
interactions of the interaction between constituent quarks are shown
in Figs.2--9 and are also listed in Table 1. The potential components
are given separately for the contribution of the two--pion exchange
loop amplitudes in Fig.2 and as obtained after the interaction
between the exchanged pions in the $S-$ and $P-$waves have been taken
into account. The calculated values at short range are very sensitive
to the choice of the value for the cut--off parameter $\Lambda$, which
here has been taken to equal the nucleon mass, and therefore these
values should be given no more than qualitative value. The sensitivity
to this cut--off is illustrated in Figs.2--9 by the curves marked
``800'', which show the results for the two-pion exchange interaction
components when 
the value of $\Lambda$ is reduced from $m_N=939$ MeV 
to 800 MeV. Several of the potential components 
change notably at very short range by that reduction, and
therefore the uncertainty range of those potential 
components for quark separations less than r $\simeq$ 0.2 fm
is large.

The general features of the calculated two--pion exchange interaction
are reminiscent of the corresponding interaction for nucleons,
although overall the interaction components are weaker in the present
case. This relative weakness is mainly due to the smallness of the
effective pion--quark coupling constant. The isospin dependent two--pion
exchange tensor and spin--spin potentials are however much stronger
than the corresponding one--pion exchange interactions in the range of
relevance for baryon structure: $0.1 \leq r \leq 0.6$ fm. Therefore
any realistic 
meson exchange model for the hyperfine interaction between
constituent quarks has to include the two--pion exchange interaction.
The effect of the two--pion exchange interaction is to in effect cancel
out the one--pion exchange tensor interaction in the relevant range,
and to strongly enhance the effect of the one--pion exchange spin--spin
interaction.

\vspace{1cm}

{\bf 5. Phenomenological considerations}

\vspace{0.5cm}

{\bf 5.1 The components of the interaction}

\vspace{0.5cm}

It is instructive to consider the two--pion exchange interaction
between constituent quarks calculated here together with the other
components of the interaction between constituent quarks that are
required for a satisfactory description of the baryon spectrum. To
these belong the confining interaction, which is the source of the
unbounded discrete spectrum, and the one--pion exchange interaction,
which appears naturally as an iteration of the instanton induced
interaction in the $t-$channel. Finally there presumably remains a
weak screened gluon exchange interaction. We propose that the
conceptually simplest phenomenologically acceptable model for the
baryon spectrum is obtained as a combination of a linear scalar
confining interaction along with the one-- and two--pion exchange
interactions and complemented by an omega exchange interaction,
which represents the most important component of the
three--pion exchange interaction. 

There are several reasons to believe that the confining interaction
should have the form of a scalar flavor independent interaction, with
leading linear component \cite{Gromes,Bali,Lahde}. That interaction
would then have the central and spin--orbit components 
$$v_{C,conf}^+(r)=cr,\quad v_{LS,conf}^+=-{c\over 2m^2 r},\eqno(5.1)$$
where $c$ is the string tension, which for quarks has the magnitude
$\sim$ 500 MeV/fm. While all realistic dynamical quark models contain
a central linear confining interaction of this form, the spin--orbit
interaction associated with the linear confining interaction has to be
cancelled by another interaction component in view of the smallness of
the spin--orbit splittings of the baryon multiplets in the $P-$shell
\cite{Carlson}.

The spin--orbit components of the two--pion exchange interaction provide
such a cancelling mechanism for 
the spin--orbit interaction in (5.1). Note first
that it is only the spin--orbit interaction in quark pairs with
antisymmetric flavor symmetry (or isospin 0), which plays a role in
the case of the $P-$shell baryons \cite{GloRis,Coester}. This implies
that the effective spin--orbit interaction in the $P-$shell is
the combination
$v_{LS}^+(r)-3v_{LS}^-(r)$. Since the spin--orbit component
$v_{LS}^-(r)$ of the two--pion exchange interaction is much stronger
than the component $v_{LS}^+(r)$ in the case of the two--pion exchange
interaction (Figs.4, 5), the combination $v_{LS}^+(r)-3v_{LS}^-(r)$
is positive. While its magnitude is very
sensitive to the model for the
$\pi\pi$ interaction in the $S-$ and $P-$waves, it is -- as
shown in Fig.10 -- in any case
much larger than that of $v_{LS,conf}^+$, for quark separations in
the relevant range, and therefore more than cancels out the effect
of the spin--orbit interaction that 
is associated with the confining interaction (5.1). 
To show the parameter sensitivity of the combination
of spin--orbit
interactions the result as obtained with a 20 \% weaker
$\rho-$meson--quark coupling constant is also shown.
The
positive net spin--orbit potential in quark pair states with 
symmetric spin and antisymmetric flavor symmetry implies 
spin--orbit splittings in the empirically indicated direction
in $P-$shell.

The spin-independent central component of the two--pion exchange
interaction is attractive in flavor antisymmetric quark pair states
and weak and repulsive in flavor symmetric states. These
interaction components should be considered in combination with the
spin--spin interaction potentials, which are much larger in magnitude.
The isospin independent and dependent spin--spin interactions combine
to a moderately strong repulsive interaction in spin 1 
and isospin 1 quark
pair states. That interaction adds to the net repulsion from the
central interaction. In spin 0 and isospin 0 quark pair states the net
interaction is in contrast strongly attractive. This 
spin and flavor
dependence of the interaction suggests that the key component of the
meson exchange hyperfine interaction is indeed a flavor--spin dependent
interaction of the form $-\vec \tau^1\cdot \vec \tau^2 \vec
\sigma^1\cdot \vec \sigma^2$. That is exactly the operator
form required for the explanation of the empirical reversal of 
normal ordering in the
baryon spectrum \cite{GloRis}. 

The isospin dependent tensor component of the two--pion exchange
interaction (Fig.7) serves to in effect cancel out the corresponding
one--pion exchange tensor interaction in the relevant range between 0.2
and 0.6 fm. This is a phenomenologically desirable feature, as (a) the
one--pion exchange tensor interaction by itself would give rise to
spin--orbit splittings of the $P-$shell baryon multiplets, which --
 while small -- 
typically go in the wrong direction and (b) as the extant evidence on
tensor interaction induced deformation of the $\Delta_{33}$ resonance
suggests that to be very small \cite{Papa}.

The situation concerning the isospin dependent spin--spin interaction
is exactly the opposite: that component of the two--pion exchange
interaction strongly enhances the corresponding component of the
one--pion exchange interaction, and explains the origin of the required
structure of the hyperfine interaction.

\vspace{1cm}

{\bf 5.2 Two-pion exchange and $\rho-$meson exchange}

\vspace{0.5cm}

The enhancement of the isospin--dependent spin--spin component of the
one--pion exchange interaction and the cancellation of much of the
tensor component is well known in nuclear physics, where nucleons are
the effective degrees of freedom. It is instructive to outline 
the reasons for this enhancement and
cancellation in nonrelativistic notation, which is generally 
sufficient
for the nuclear case. The mechanisms are the same as those operating
between constituent quarks.

The nuclear one--pion exchange interaction, that corresponds 
to $K_\pi$ in eq. (2.3), is

$$V_\pi(\vec k)=-{f^2\over m_\pi^2}
(\vec \tau_1\cdot \vec \tau_2){(\vec
\sigma_1\cdot \vec k)( \vec\sigma_2\cdot \vec k)
\over k^2+m_\pi^2}\eqno(5.2)$$
where retardation has been neglected. The 
interaction $V_\pi(\vec k)$ can be
separated into spin--spin and tensor components as:
$$V_\pi(\vec k)=-{1\over 3}{f^2\over m_\pi^2}(\vec \tau_1\cdot \vec
\tau_2)(\vec \sigma_1\cdot \vec \sigma_2)\{1-{m_\pi^2\over
k^2+m_\pi^2}\}$$
$$-{f^2\over m_\pi^2}(\vec \tau_1\cdot \vec \tau_2)\{{\vec
\sigma_1\cdot \vec k\, \vec \sigma_2\cdot \vec k-{1\over 3}\vec
\sigma_1\cdot \vec \sigma_2 k^2\over k^2+m_\pi^2}\},\eqno(5.3)$$
where we have rewritten $k^2/(k^2+m_\pi^2)$ as
$1-m_\pi^2/(k^2+m_\pi^2)$ in the first term.

The term with the 1 in the curly brackets 
upon transformation to configuration space
represents a zero--range
interaction
which will be
strongly modified in the presence of short--range repulsive
interaction resulting,
e.g., from vector--meson exchange, which keep the interacting
particles apart.

The spin-- and isospin--dependent part of the two--pion exchange
interaction consists, as noted above, of a strongly correlated
system of pions in a relative $P-$state, essentially a $\rho-$meson of
distributed mass. The form of such an interaction between nucleons
is obtained from the 
transverse coupling of the $\rho-$meson to the nucleon:
$${\cal L}_\rho=f_\rho\bar\psi(x)(\vec \sigma\times
\nabla)\cdot \vec \rho \cdot \vec \tau \psi(x).\eqno(5.4)$$
The nucleon--nucleon -- or quark--quark -- interactions resulting 
from $\rho-$exchange with
this tensor coupling is
$$V_\rho(\vec k)=-{f_\rho^2\over m_\rho^2}(\vec \tau_1\cdot \vec
\tau_2){(\vec \sigma_1\times \vec k)\cdot 
(\vec \sigma_2\times \vec
k)\over k^2+m_\pi^2}.\eqno(5.5)$$
Rearrangement into spin--spin and tensor components yields
$$V_\rho(\vec k)=-{2\over 3}{f_\rho^2\over m_\rho^2}(\vec \tau_1\cdot
\vec \tau_2)(\vec\sigma_1\cdot \vec \sigma_2)\{1-{m_\rho^2\over
k^2+m_\rho^2}\}$$
$$+{f_\rho^2\over m_\rho^2}(\vec \tau_1\cdot \vec \tau_2)\{{(\vec
\sigma_1\cdot \vec k)(\vec \sigma_2\cdot \vec k)-{1\over 3}(\vec
\sigma_1\cdot \vec \sigma_2)k^2\over k^2+m_\rho^2}\}.\eqno(5.6)$$
Comparison of eq.(5.6) with eq.(5.3) reveals that the spin--spin
terms from $\pi-$ and $\rho-$ exchange have the same sign, whereas the
tensor terms have opposite signs. This argument concerning the
relative signs carries over directly to the case of
the quark-quark interaction.

Although the $\rho-$meson couples transversely to the spin in eq.
(5.4), modification of the zero--range part of the interaction to take
into account short--range interactions, as discussed following eq.
(5.3), introduces an effective longitudinal 
coupling which turns out to be
important in pionic excitations in nuclei.

\vspace{1cm}

{\bf 5.3 $\omega$ meson and gluon exchange}

\vspace{0.5cm}

The isospin independent tensor component of the two--pion exchange
interaction (Fig.6) is strong and attractive. Since the baryon
spectrum reveals no phenomenological indication for such an
interaction, this interaction component
has to be cancelled by a tensor potential of opposite
sign from another mechanism. In view of the significance of the
$\rho-$meson resonance component of the two--pion exchange interaction
it is natural to invoke an $\omega$ meson exchange mechanism between
quarks to bring about this cancellation, as that has about the same
range and strength as the $\rho-$meson exchange interaction.

The $\omega$ meson exchange mechanism forms the most significant
component of the three--pion exchange interaction. This is described by
the following potential components:
$$v_{C,\omega}^+(r)=m_\omega{g_{\omega qq}^2\over 4\pi}
{e^{-m_\omega r}\over
m_\omega r},$$
$$v_{LS,\omega}^+(r)=-{3m_\omega\over 2}{g_{\omega qq}^2\over
4\pi}({m_\omega\over m_q})^2(1+{1\over m_\omega r}){e^{-m_\omega
r}\over m_\omega^2r^2},$$
$$v_{T,\omega}^+(r)=-{m_\omega\over 12}{g_{\omega qq}^2\over
4\pi}({m_\omega\over m_q})^2({3\over m_\omega^2r^2}+{3\over m_\omega
r}+1){e^{-m_\omega r}\over m_\omega r},$$
$$v_{SS,\omega}(r)={m_\omega\over 6}{g_{\omega qq}^2\over
4\pi}({m_\omega\over m_q})^2\{{e^{-m_\omega r}\over m_\omega
r}-{4\pi\over m_\omega^3}\delta^{(3)}(r)\}.\eqno(5.7)$$
Here $m_\omega$ is the $\omega-$meson mass (783 MeV) and $g_{\omega
qq}$ is the $\omega-$quark vector coupling constant. In the presence
of a monopole form factor for the $\omega-N$ interaction, the terms on
the r.h.s. should be amended by corresponding terms with opposite sign
and with $m_\omega$ replaced by the factor mass scale $\Lambda$.

The quark model relation between the $\omega-$quark and
$\omega-$nucleon coupling constants is $g_{\omega qq}=g_{\omega NN}/3$
in the case of the spin--independent part of the coupling
and $g_{\omega qq}=(m_q/m_N)g_{\omega NN}$ in the case of the
spin--dependent part of the coupling. As $m_q/m_N$ is close
to 1/3 we shall use the former relation here. With
$g_{\omega NN}^2/4\pi\simeq 20$ \cite{Machleidt} we then
have $g_{\omega qq}$=5.3.

The components of the $\omega-$meson exchange interaction between
constituent quarks are shown in Fig.11, as obtained with 
$g_{\omega qq}=5.3$ and with the cut off mass scale of $\Lambda=m_N$
also used in the calculation of the two--pion exchange potential
above. The $\omega$ exchange potential has the same functional
form as a screened gluon exchange interaction, but with opposite
overall sign. This makes it difficult to separate phenomenologically
the effects of omega and gluon exchange on the baryon spectrum,
as a stronger $\omega-$quark coupling may be compensated by a
correspondingly stronger effective quark--gluon coupling. The
expression for a screened one-gluon exchange interaction may be
obtained from the expressions (5.2) for the omega exchange
interaction by the replacements
$${g_{\omega qq}^2\over 4\pi}\rightarrow -{2\over 3}\alpha_S,
\quad m_\omega\rightarrow m_G.     \eqno (5.8)$$
Here $\alpha_S$ is the effective color hyperfine constant,
and $m_G$ is a screening mass for the gluon exchange
interaction. The screening mass should fall somewhere between
$\Lambda_{QCD}\sim 250$ MeV and $\Lambda_\chi$ \cite{HelRis}.

The $\omega-$exchange potential components shown in Fig.11
should be combined with the isospin independent two--pion
exchange interaction components in Figs.2--9. The central
component of the $\omega-$exchange interaction cancels
most of the corresponding component of the two--pion
exchange interaction (Fig.2). This suggests that the
only significant spin-- and isospin--independent interaction
between two constituent quarks is the central component of
the confining interaction, as shown in Fig.10,
where the combined meson exchange and confinement contributions
to the isospin independent central interaction are shown.

The spin--orbit component of the $\omega-$exchange 
interaction adds to that generated by two--pion
exchange (Fig.4), but it brings no qualitative change.
That interaction, while strong, is overwhelmed by the
contribution of the isospin dependent two--pion
exchange spin--orbit interaction in the flavor antisymmetric
quark pair states, which determine the spin--orbit
splitting of the $P-$shell baryon resonances. This is
shown in Fig.10, where the net spin--orbit interaction
for such pair states is shown.
Because the net spin--orbit interactions in the 
$P-$shell is weighted by a factor $r^2$ in the
matrix elements, there is a strong cancellation between
the long range spin-orbit interaction associated with
the confining interaction combined with that of $\omega-$exchange 
and that due to two-pion
exchange. 

The $\omega-$exchange tensor interaction adds to the
isospin independent part of the two--pion exchange
tensor interaction (Fig.6), but it is weaker in
magnitude. As the net isospin dependent tensor
interaction is very weak (Fig.7), the isospin
independent tensor interaction is the main tensor
component of the mesonic part of the
hyperfine interaction between
quarks. This tensor component has opposite sign to
that of one-gluon exchange or one--pion exchange
in isospin symmetric quark pair states, but it
is comparable in magnitude, and has the same sign
as the one--pion exchange tensor interaction in
isospin antisymmetric quark pair states.
The strong isospin dependent tensor interaction,
which is associated with irreducible $\pi-$gluon
exchange \cite{HelRis}, even in the case of a
weak screened quark--gluon coupling, would however
cancel out most of this tensor interaction
in the flavor antisymmetric quark pair states, which
determine the spin--orbit splitting of the
$P-$shell baryon multiplets.

Finally the spin--spin component of the $\omega$ exchange 
interaction has the opposite sign to the
corresponding component of the two--pion exchange
interaction (Fig.8). The role of $\omega$ exchange
is therefore to reduce the strength of the isospin
independent spin--spin interaction. This supports
the role of the isospin dependent one-- and two--pion
exchange spin--spin potential as the dominant
source of hyperfine splitting between constituent
quarks.

\vspace{1cm}

\centerline{\bf 6. Discussion}

\vspace{0.5cm}

The instanton liquid model of the QCD vacuum, which
appears to be supported by comprehensive lattice 
calculations \cite{Negele1,Negele2}, 
implies pointlike interactions between
quarks. To obtain a realistic description of the
hyperfine interaction between constituent quarks,
that interaction has to be iterated (at least once)
in the $t-$channel, to overcome the restriction to
flavor--antisymmetric quark pairs. This is a necessary
requirement, as the hyperfine interaction has to
have about the same strength in the (completely)
flavor symmetric $\Delta-$spectrum as in the
nucleon spectrum. The $t-$channel iteration of the
pointlike instanton induced interaction has an 
obvious meson exchange interpretation. In the
pseudoscalar channel the iteration corresponds to
Goldstone boson exchange with long range -- i.e.
the pion exchange interaction. Once pion exchange
is operative, with a relatively strong coupling
strength, it follows that two--pion exchange 
also plays a significant role, as indeed found in
the present calculation.

Phenomenological analysis of the baryon spectrum reveals
that it is simplest to describe in the constituent
quark model with a central linear
confining interaction in combination with an attractive
flavor-spin hyperfine interaction \cite{GloRis,Hollis}. 
The present
results show that the latter has a strong two-pion
exchange contribution at scales commensurate with
the range of the baryon wave functions, if its longest
range term is due to one-pion pion exchange. The 
isospin dependent tensor
components of the one-- and two--pion exchange interactions
largely cancel, which is a phenomenologically desirable
feature. Moreover the combination of the spin-orbit
components of the two-pion exchange interactions that
appears in the $P-$shell multiplets is strong and
overwhelms the long range spin-orbit interaction that
would be associated with a linear scalar confining
interaction. These qualitative conclusions are not
hinged on the strength of the interaction between
the exchanged pions in the $S-$ wave -- i.e. whether
it is resonant or the strength of the coupling of
that resonance to quarks.

The meson exchange description of the hyperfine
interaction between constituent quarks has an 
obvious similarity to the meson exchange description
of the nucleon--nucleon interaction. The similarity is
not complete, however, as in the case of constituent
quarks, the coupling to mesons should disappear beyond
the chiral restoration scale $\Lambda_\chi$. This
implies that the meson exchange interaction should
vanish at ranges  shorter than $\Lambda_\chi^{-1}$. Consequently
only few--pion exchange mechanisms need to be invoked.
Here we have considered one-- and two--pion exchange
and have approximated the main three--pion exchange
interaction by $\omega-$meson exchange. Uncorrelated
three--pion exchange should be weak, and may in
principle be calculated by the methods in ref.\cite{Hamilton}.  

Another notable difference between the two--pion exchange
interactions between nucleons and quarks follows from
the absence of excited states of quarks. The two--pion
exchange interaction between quarks therefore has
no analogue to the strongly atteractive terms that
arises from loop diagrams with intermediate nucleon
resonances -- first and foremost the $\Delta_{33}$ --
which provide the bulk of the attraction between
nucleons.

In the present calculation of the two--pion exchange interaction
the constituent quarks have been treated as Dirac particles,
with constant mass. As the effect of confinement is not
taken into account in the quark propagators, the results
should be viewed as qualitative and suggestive rather
than quantitative. The confining interaction may in
principle be taken into account by treating the constituent
mass as a running mass, which grows linearly with separation
of the quarks. At the length scales relevant for the
structure of the baryons $\sim$ 0.2--0.6 fm, the treatment
of the constituent quark mass as an average constant
should not be expected as very unrealistic, as all
quark model descriptions of the baryon spectrum are
based on that approximation.

\vspace{1cm}

\centerline{\bf Acknowledgments}

\vspace{0.5cm}

We are indebted to Professor R. McKeown and W. Haxton for 
hospitality
at the W. K. Kellogg Radiation Laboratory at the California
Institute for Technology and the Institute for Nuclear Theory
at the University of Washington respectively for hospitality
during the completion of this work. 
DOR thanks Dr. L. Ya. Glozman for instructive
corresondence.
Research supported in part
by the Academy of Finland by grants No. 34081, 43982 and
the U. S. Department of Energy under grant 
DE-FG02-88ER40388.

\newpage

\centerline{\bf Table 1}

\vspace{1cm}

\begin{center}
\begin{tabular}{|l|l|l|l|l|l|l|l|l|}\hline
$r$(fm) & $v_C^+$ & $v_C^-$ & $v_{LS}^+$ & $v_{LS}^-$ & $v_T^+$ &
$v_T^-$ & $v_{SS}^+$ & $v_{SS}^-$\\\hline
0.1 & -420 & 1800 & -88700& -64300 & -24800& 3930 & 7380 & -4180\\
0.2 & -201 & 655 & -7270 & -9360& -2100 & -80.6 & 1060 &
-458\\
0.3 & -108 & 317 & -1477 & -2740 & -422 & -122 & 276 &
-68.7\\
0.4 & -61.4 & 172 & -439 & -1050 & -121 & -72.4 & 92.7 & 0.14\\
0.5 & -36.6 & 99.9 &-161 & -465 & -41.9 & -40.9 & 35.8 & 11.7\\
0.6 & -22.6 & 59.1 & -68.4 & -224 & -16.6 & -23.3 & 15.2 & 11.1\\
0.7 & -14.4 & 36.2 & -32.0 & -115 &-7.25 & -13.6 & 6.96 & 8.42\\
0.8 & -9.4 & 22.6 & -16.2 & -61.4 & -3.41 & -8.07 & 3.38 & 5.90\\\hline
\end{tabular}
\end{center}
\vspace{0.5cm}

Table 1. The components (in MeV) of the two--pion exchange interaction
between quarks as defined in (4.3). The numerical values correspond to
the case, when both the $S-$ and $P-$wave interactions between the
exchanged pions have been taken into account.

\vspace{1cm}

\centerline{\bf Figure Captions }

\vspace{1cm}

Fig.1 Two--pion exchange loop amplitudes. The fermion
lines represent $u$ and $d$ quarks.

Fig. 2 Isospin independent central interaction $v_C^+$. The curve
``box'' is the result obtained from the two-pion exchange loop
diagrams in Fig.1, the curve ``S-wave interaction'' is the
result obtained after the interaction in the $S-$state
of the $\pi\pi$ system is taken into account. The curve ``800''
shows the latter result when the cut--of mass is taken to
be 800 MeV instead as $m_N$.

Fig. 3   Isospin dependent central interaction $v_C^-$. The
curve ``box'' is the result obtained from the two-pion exchange loop
diagrams in Fig.1, the curve ``P-wave interaction'' is the
result obtained after the interaction in the $P-$state
of the $\pi\pi$ system is taken into account. The curve ``800''
shows the latter result when the cut--of mass is taken to
be 800 MeV instead as $m_N$.

Fig. 4 Isospin independent spin--orbit interaction $v_{LS}^+$. The curve
``box'' is the result obtained from the two-pion exchange loop
diagrams in Fig.1, the curve ``S-wave interaction'' is the
result obtained after the interaction in the $S-$state
of the $\pi\pi$ system is taken into account. The curve ``800''
shows the latter result when the cut--of mass is taken to
be 800 MeV instead as $m_N$.

Fig. 5   Isospin dependent spin--orbit interaction $v_{LS}^-$. The
curve ``box'' is the result obtained from the two-pion exchange loop
diagrams in Fig.1, the curve ``P-wave interaction'' is the
result obtained after the interaction in the $P-$state
of the $\pi\pi$ system is taken into account. The curve ``800''
shows the latter result when the cut--of mass is taken to
be 800 MeV instead as $m_N$.

Fig. 6 Isospin independent tensor interaction $v_T^+$. The curve
``box'' is the result obtained from the two-pion exchange loop
diagrams in Fig.1 and the curve ``800''
shows the result when the cut--of mass is taken to
be 800 MeV instead as $m_N$.

Fig. 7   Isospin dependent tensor interaction $v_T^-$. The
curve ``box'' is the result obtained from the two-pion exchange loop
diagrams in Fig.1, the curve ``P-wave interaction'' is the
result obtained after the interaction in the $P-$state
of the $\pi\pi$ system is taken into account. The curve ``800''
shows the latter result when the cut--of mass is taken to
be 800 MeV instead as $m_N$ and the curve ``OPEP'' is the
one-pion exchange component.

Fig. 8 Isospin independent spin--spin interaction $v_{SS}^+$. The curve
``box'' is the result obtained from the two-pion exchange loop
diagrams in Fig.1 and the curve ``800''
shows the result when the cut--of mass is taken to
be 800 MeV instead as $m_N$.

Fig. 9   Isospin dependent spin--spin interaction $v_{SS}^+$. The
curve ``box'' is the result obtained from the two-pion exchange loop
diagrams in Fig.1, the curve ``P-wave interaction'' is the
result obtained after the interaction in the $P-$state
of the $\pi\pi$ system is taken into account. The curve ``800''
shows the latter result when the cut--of mass is taken to
be 800 MeV instead as $m_N$ and the curve ``OPEP'' is the
one-pion exchange component.

Fig. 10 The contributions to the spin--orbit interaction for quark
pairs with antisymmetric flavor symmetry, which is the active
part of the spin--orbit interaction in the $P-$shell baryons.
The curve TWOPI1 represents the combination $v_{LS}^+ -3v_{LS}^-$
of the two isospin components of the two--pion exchange interaction
while the curve TWOPI2 gives the corresponding results when
the $\rho-$meson-quark coupling has been reduced by 20 \%.
The
curve CONF is the spin--orbit component of the confining interaction,
and the curve $OMEGA$ is the spin-orbit component of the 
$\omega-$meson exchange interaction.

Fig. 11 The components of the $\omega-$meson exchange interactions
between constituent quarks. The curves C, LS, T and SS represent 
$v_{C,\omega}^+$, $v_{LS,\omega}^+$, $v_{T,\omega}^+$ and
$v_{SS,\omega}$ respectively.

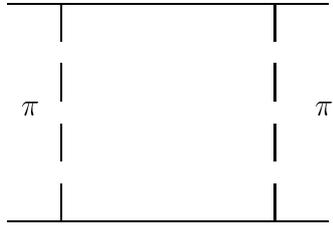
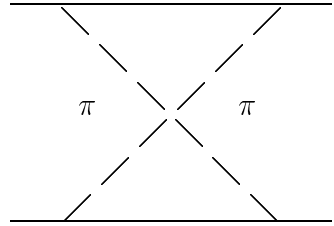
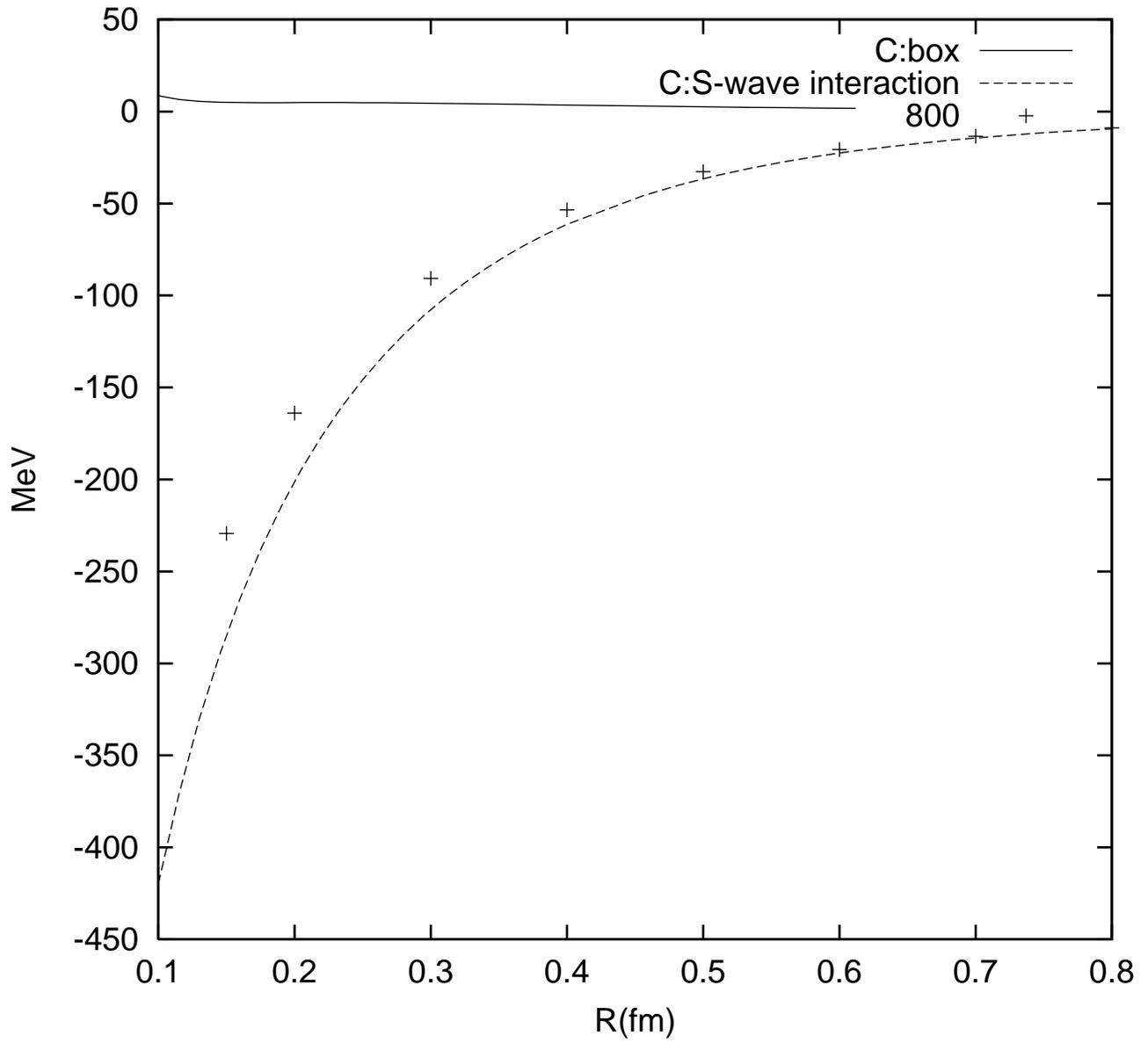
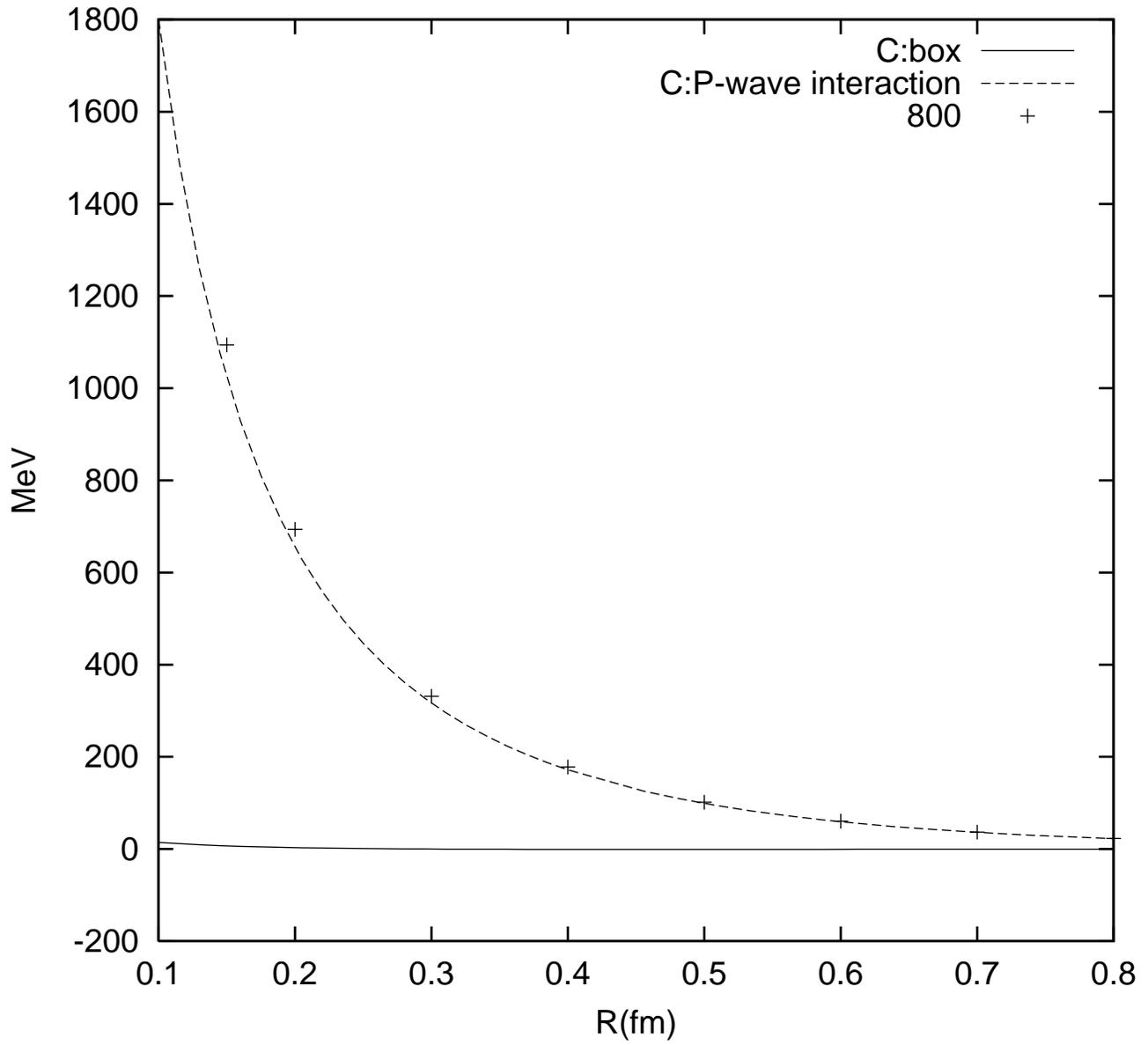
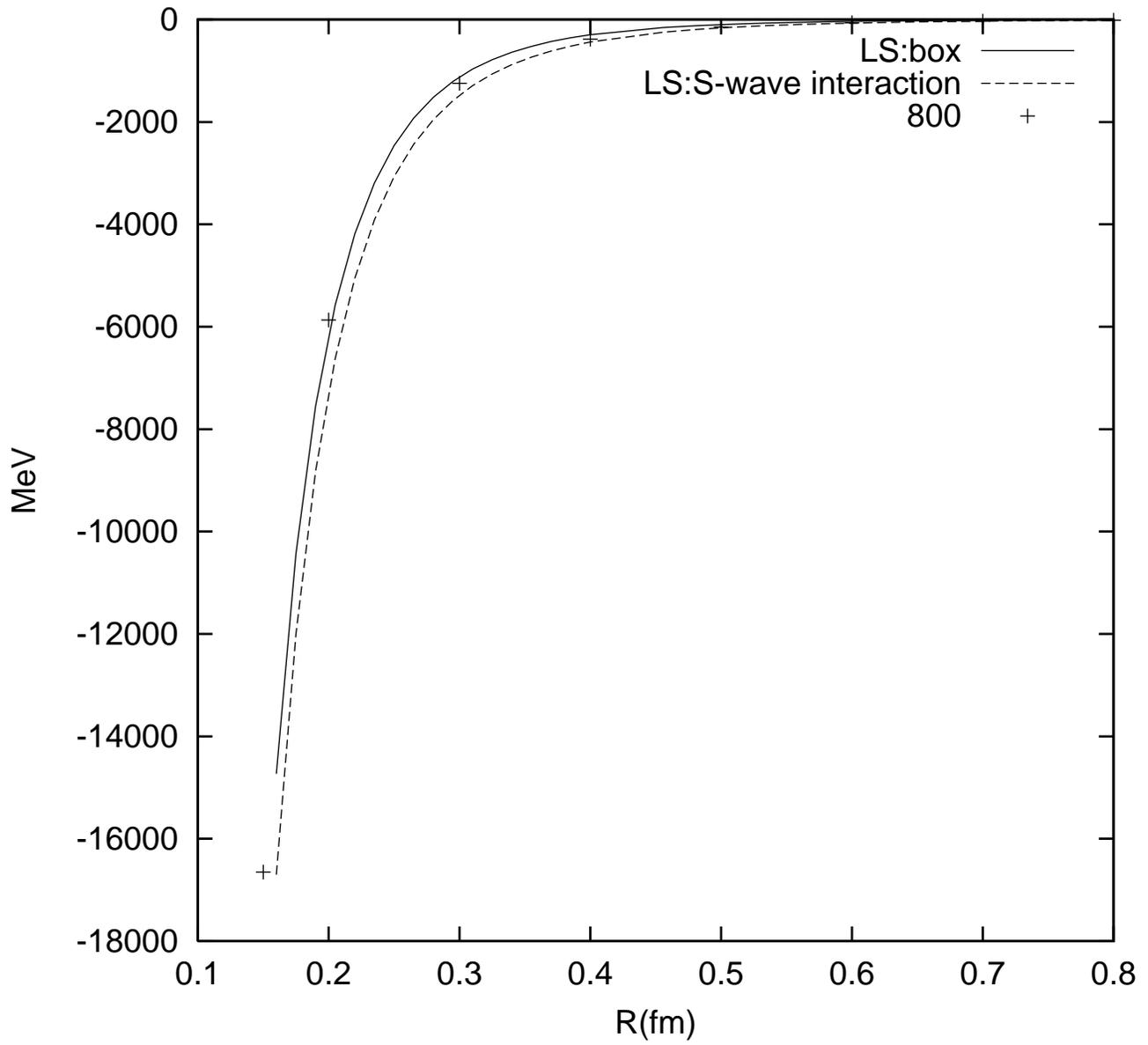
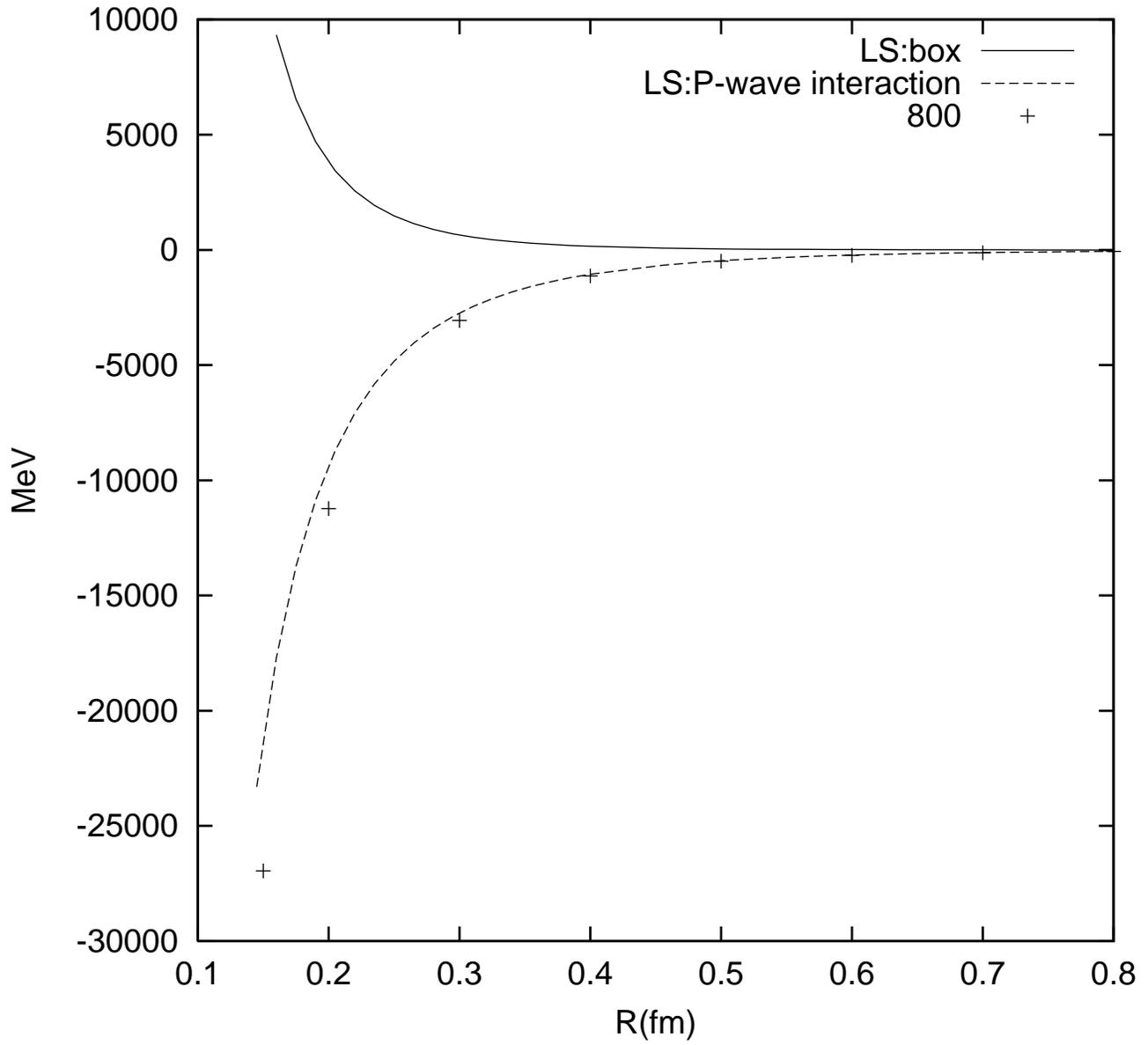
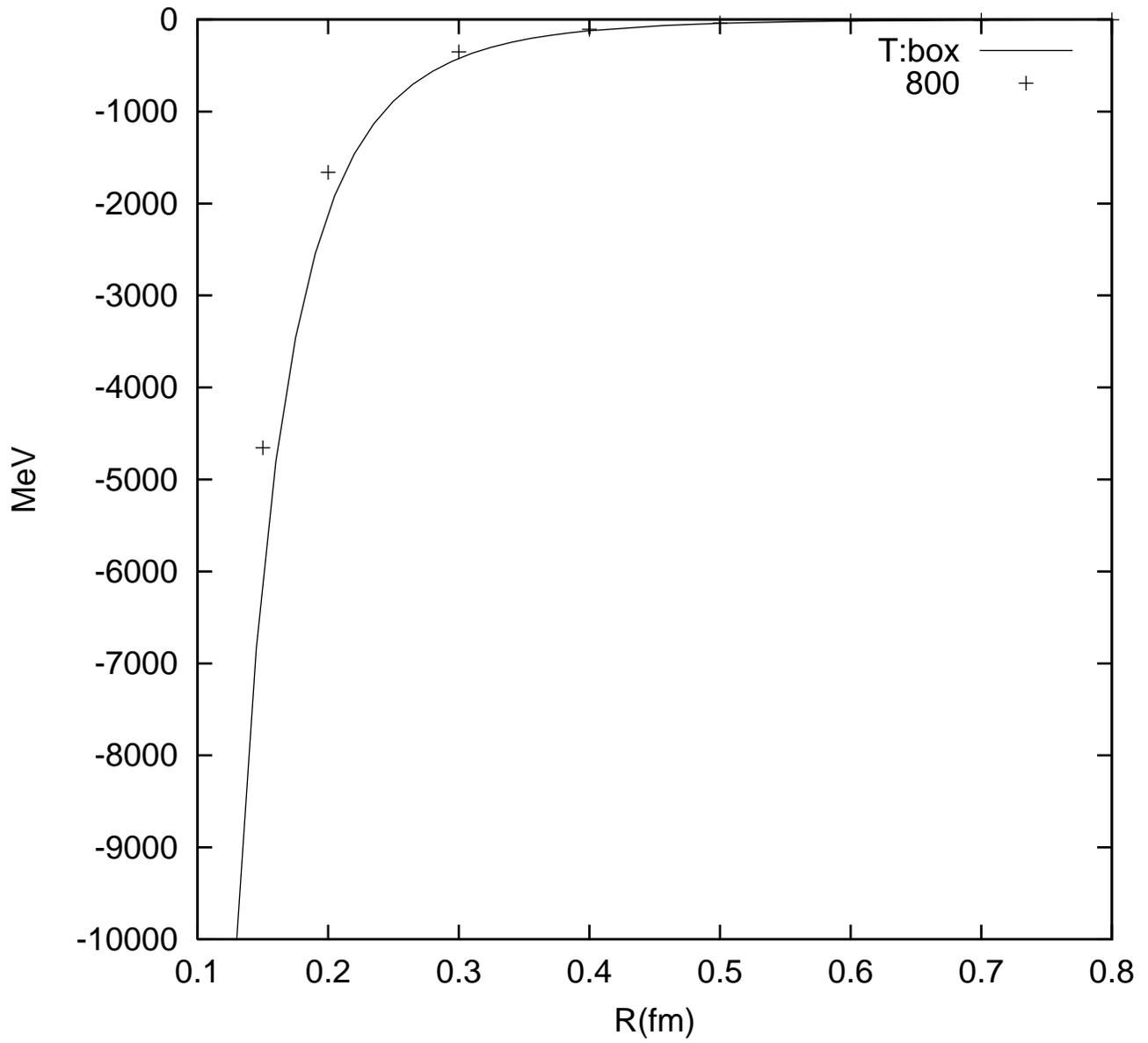
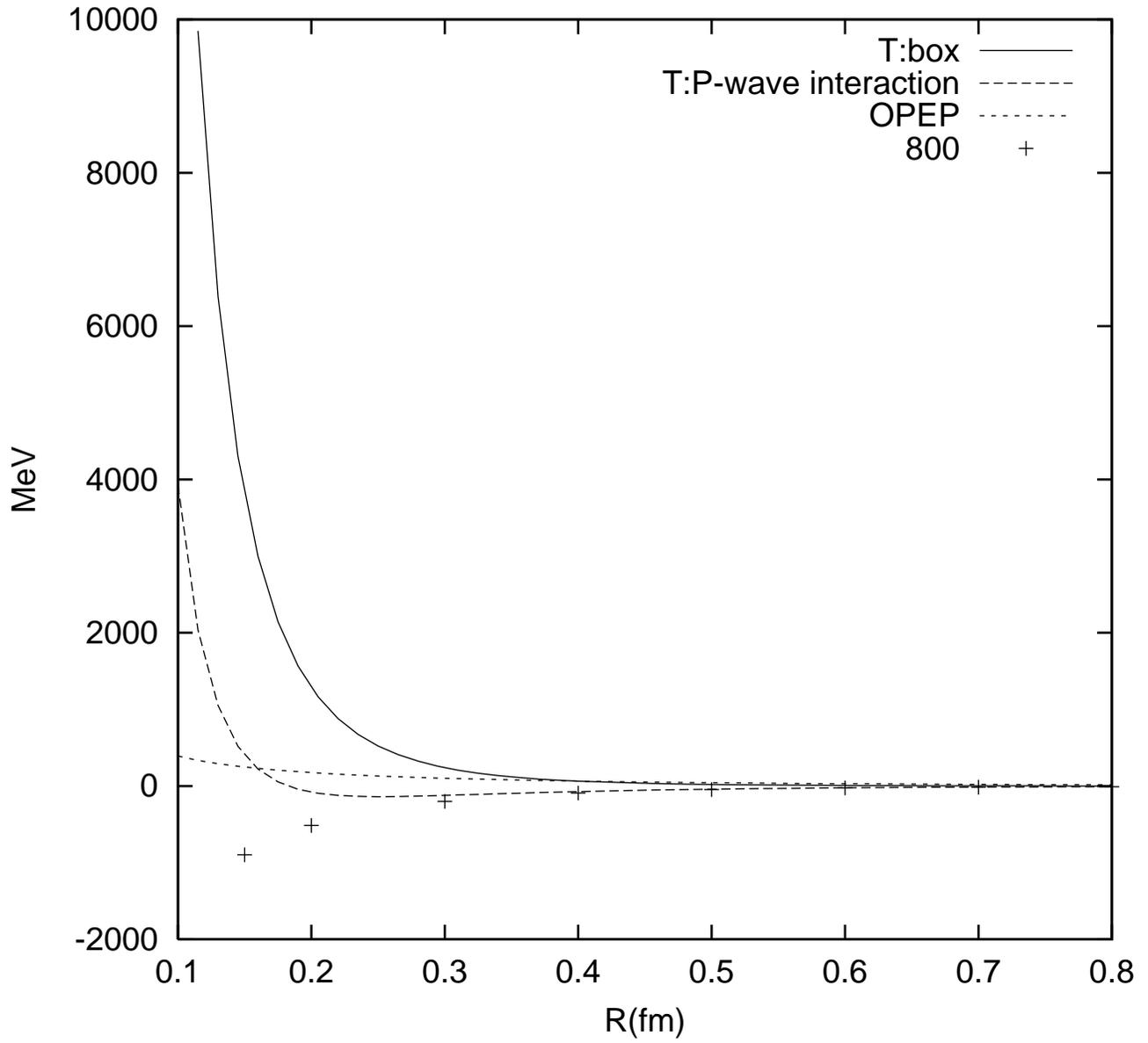
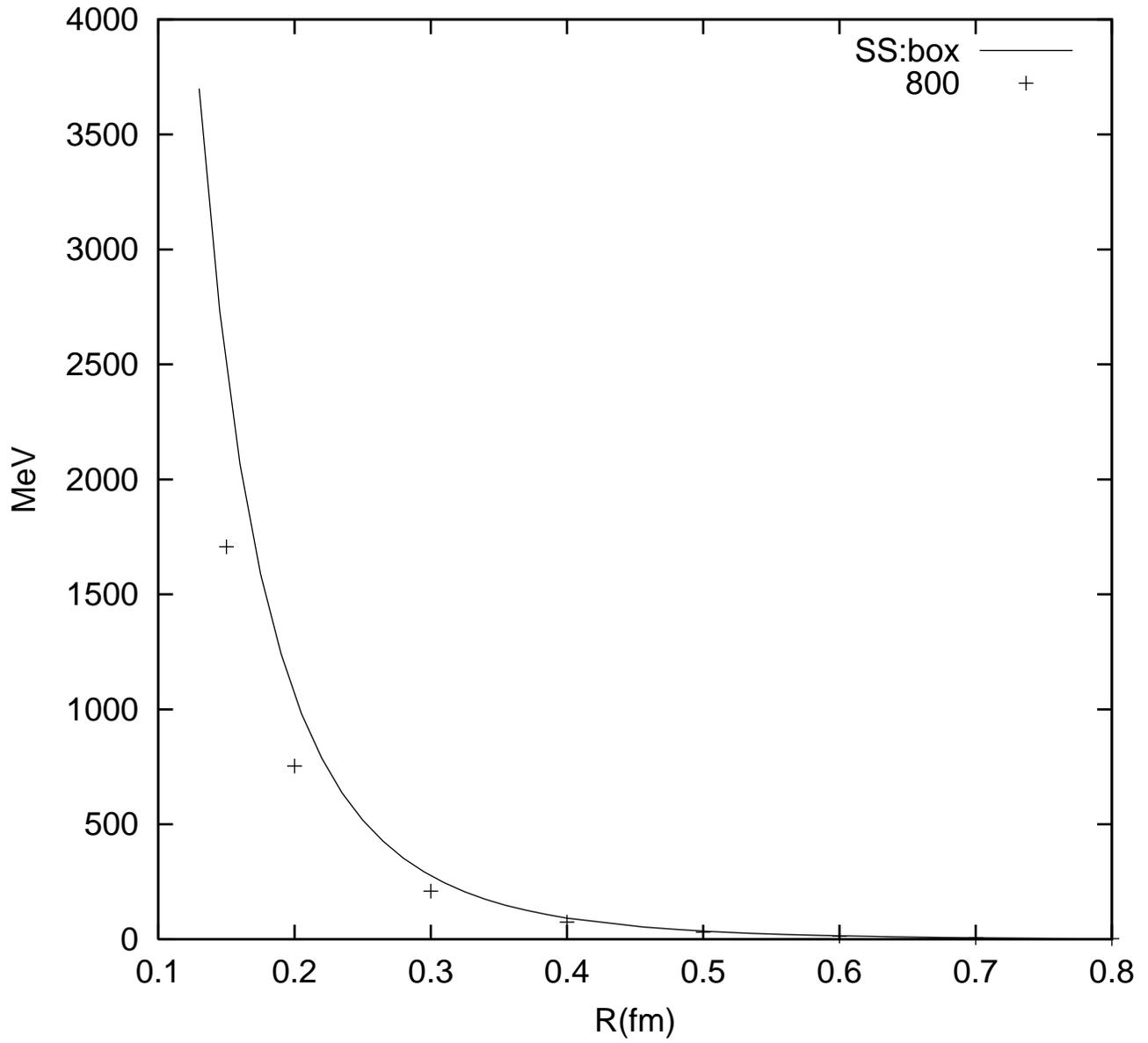
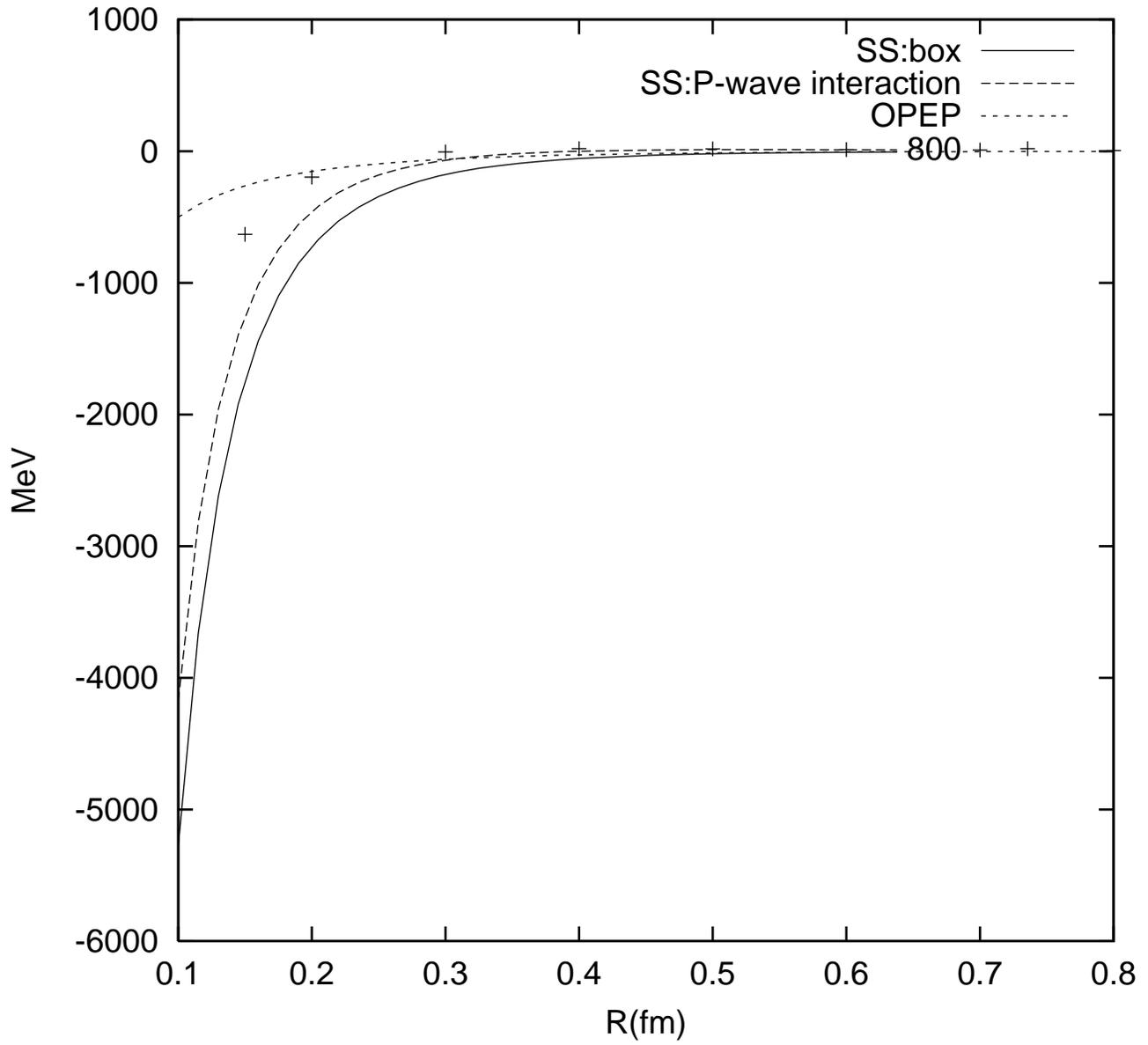
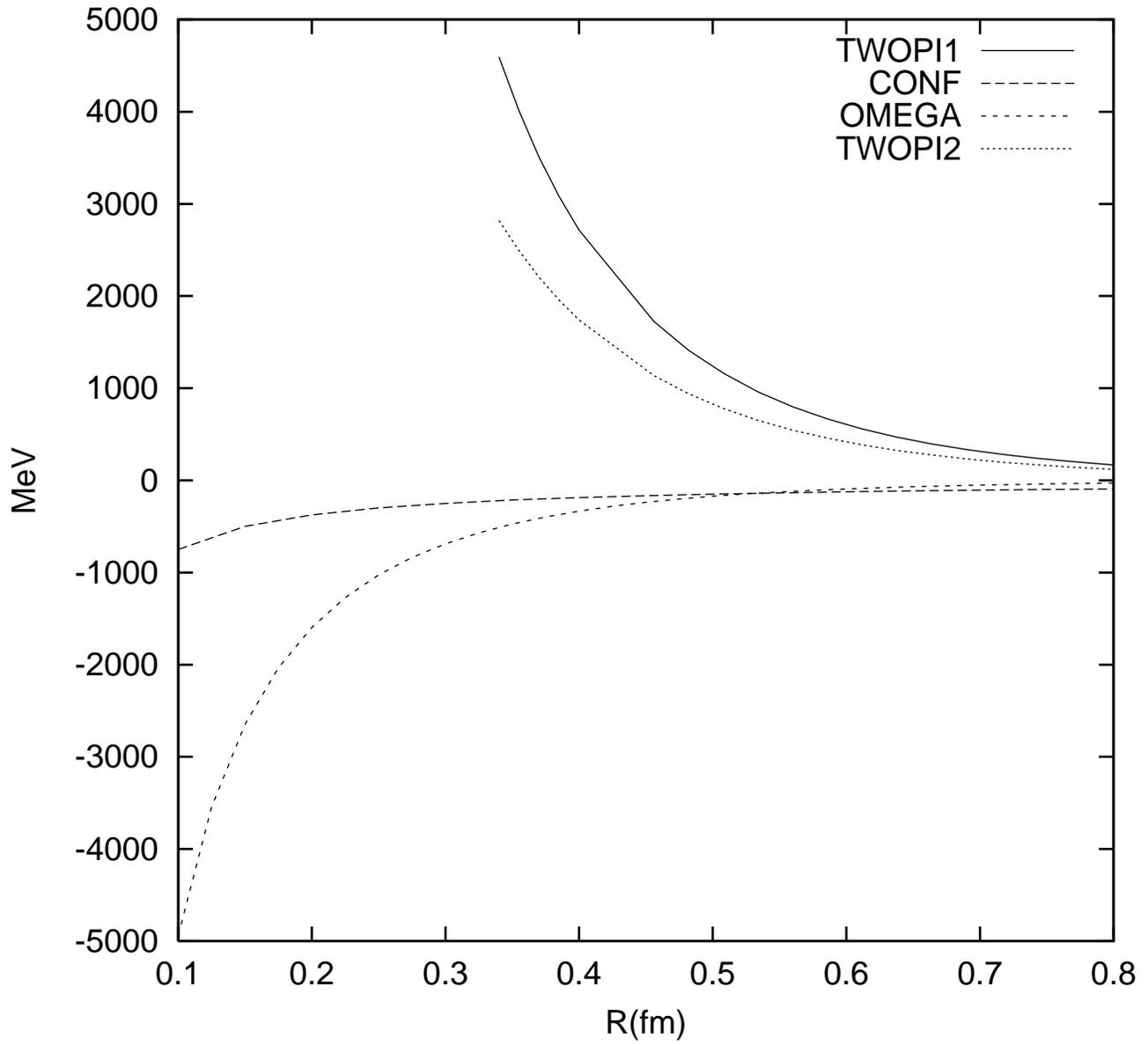
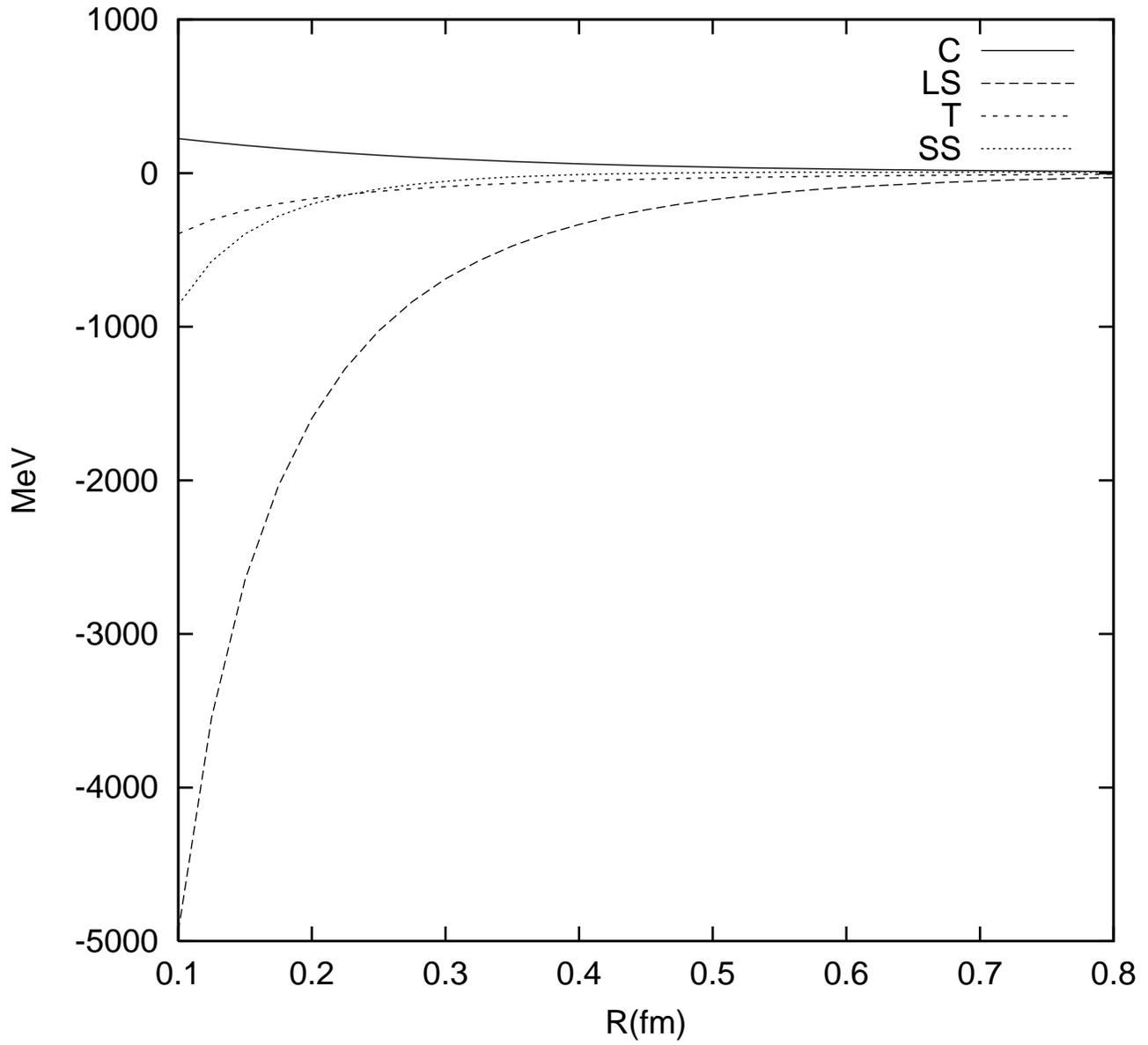
\end{document}